# Full band Monte Carlo simulation of AlInAsSb digital alloys

Jiyuan Zheng,[1] Sheikh Z. Ahmed,[1,2] Yuan Yuan,[1] Andrew Jones,[1] Yaohua Tan,[3,a] Ann K. Rockwell,[4] Stephen D. March,[4] Seth R. Bank,[4] Avik W. Ghosh,[1,2] and Joe C. Campbell[1,b]

[1]*Electrical and Computer Engineering Department, University of Virginia, Charlottesville, Virginia 22904, USA*

[2]*Department of Physics, University of Virginia, Charlottesville, Virginia 22904, USA*

[3]*Synopsys Inc, 455 N Mary Ave, Sunnyvale, CA 94085, USA*

[4]*Microelectronics Research Center, University of Texas, Austin, Texas 78758, USA*

Avalanche photodiodes fabricated from AlInAsSb grown as a digital alloy exhibit low excess noise. In this paper we investigate the band structure-related mechanisms that influence impact ionization. Band-structures calculated using an empirical tight-binding method and Monte Carlo simulations reveal that the mini-gaps in the conduction band do not inhibit electron impact ionization. Good agreement between the full band Monte Carlo simulations and measured noise characteristics is demonstrated.

**Keywords:** AlInAsSb, digital alloy, avalanche photodiode, First principle study

## I. INTRODUCTION

For optical communications and data transmission applications, high-sensitivity receivers can reduce energy consumption and system cost. The internal gain of avalanche photodiodes (APDs) can provide a sensitivity advantage relative to p-i-n photodiodes. The excess noise factor of an APD, which arises from the random nature of impact ionization, is a key factor in the receiver sensitivity. The excess noise power density can be expressed as $\Phi = 2qIR(\omega)M^2F(M)$,[1] where $q$, $I$, and $R$ represent electron charge, current, and device impedance, respectively. In the local-field model, $F(M) = kM + (1 - k)(1 - 1/M)$, where $k$ is the ratio of the hole ionization coefficient, $\beta$, to that of the electron, $\alpha$.[1] The excess noise increases with gain but increases more slowly for low k values. The highest APD receiver sensitivities in the fiber optic 1310 nm and 1550 nm transmission windows have been achieved by InGaAs/InAlAs and Ge/Si separated absorption, charge, and multiplication (SACM) structures. The performance of InGaAs/InAlAs SACM APDs is limited by their relatively high k value, 0.2. While the k value of Ge/Si APDs is an order of magnitude lower, the large lattice mismatch between Ge and Si gives rise to large dark current, which yields noise comparable to InGaAs/InAlAs. Recently, an AlInAsSb digital alloy-based APD has been demonstrated to have low $k$ (~0.01) and low dark current. In this paper, we use an empirical tight-binding method to calculate the band structure and Monte Carlo simulations to analyze the physical mechanisms responsible for low noise in this material system.

---

[a] This research was performed while Y. Tan was at the Electrical and Computer Engineering Department, University of Virginia, Charlottesville, Virginia 22904, USA.

[b] Corresponding author: Joe C. Campbell. Electronic mail: jcc7s@eservices.virginia.edu

Random alloys of $Al_xIn_{1-x}As_ySb_{1-y}$ with a high Al content exhibit a wide miscibility gap A recent set of experiments and calculations on APDs have uncovered some highly promising design possibilities hitherto unexplored in the field. This involves the ability by the Banks group at UT to grow certain digital alloys, such as $Al_xIn_{1-x}As_ySb_{1-y}$ with a high Al content, which in random alloys exhibit a wide miscibility gap.[2,3] Recently it has been reported that high-Al-content $Al_xIn_{1-x}As_ySb_{1-y}$ can be grown within the miscibility gap by MBE as a digital alloy of the component binaries, AlAs, AlSb, InAs, and InSb.[4] In this work, we consider 70% Al composition. As shown in Figure 1(A), each period consists of 3 monolayers (ML) InAs, 3 ML AlSb, 1 ML AlAs and 3 ML AlSb. The 3-D bandstructure has been calculated using an empirical tight-binding model.[5-7] The model is environmentally dependent, which means the strain and the interface issue have been considered in the calculation. The empirical parameter was adjusted iteratively to fit empirical tight-binding results with a hybrid functional band structure. In this work, we used the environment-dependent sp3d5s* tight binding-model introduced in Ref 5. The bandstructure from the Γ point along the [001] direction is shown in Figure 1(B). The bandgap is 1.19 eV. A large minigap is observed in the conduction band. Intuitively, a minigap in the conduction band would prevent electrons from achieving sufficient energy to impact ionize. However, previous work has shown that electron impact ionization occurs at a much higher rate than that of holes in the AlInAsSb digital alloy.[8] In this paper, full band structure based Monte Carlo simulation is used to analyze carrier transport.[9-12]

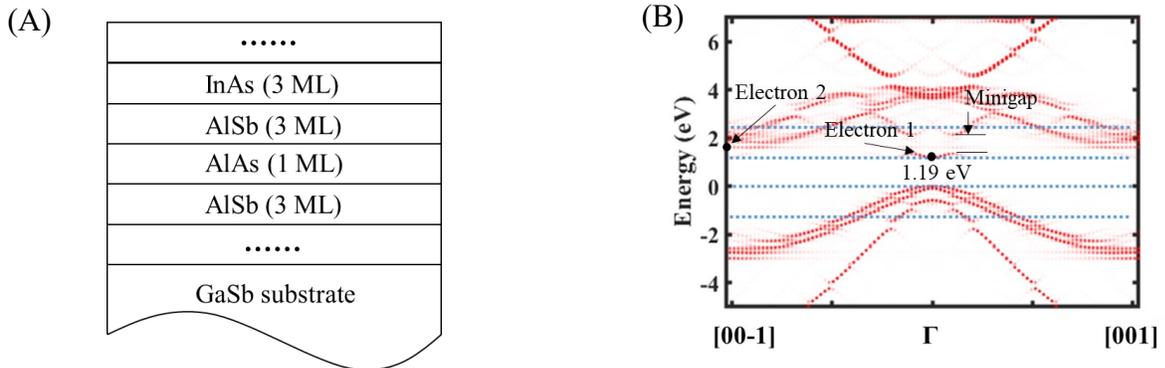

**FIGURE 1** A, Lattice structures for AlInAsSb digital alloy and its B, bandstructure and the two types of initial states of electrons in Monte Carlo simulation.

The first Brillouin zone is divided into 3-D meshes (19×19×19); 46 conduction bands and 24 valence bands are included. Deformation scattering and impact ionization are included in the Monte Carlo simulation. The carrier is considered to be in free flight between scattering events. The 3D based deformation scattering rate $P_{vv',\eta}^{def}(k, \Omega_{k\pm q})$ from a point $k$ in band $v$ to a region $\Omega_{k'}$, in band $v'$ centered around $k'$ can be derived from the Fermi Golden Rule and expressed as[9,10]

$$P_{\nu\nu',\eta}^{def}(\boldsymbol{k},\Omega_{\boldsymbol{k}\pm\boldsymbol{q}}) = \frac{\pi}{\rho\omega_{\eta q}}|\Delta^{\eta}(\nu',\boldsymbol{k},\boldsymbol{q},\nu)|^2|I(\nu,\nu';\boldsymbol{k},\boldsymbol{k}\pm\boldsymbol{q})|^2 D_{\nu'}(E',\Omega_{\boldsymbol{k}'})\left(N_{\eta q}+\frac{1}{2}\mp\frac{1}{2}\right) \quad (1)$$

where $\rho$ is the lattice density, $\boldsymbol{q}$ is the phonon wave vector of mode $\eta$, and $\Delta^{\eta}(\nu',\boldsymbol{k},\boldsymbol{q},\nu)$ is the deformation potential, which is taken to be 0.73 eV/Å in the conduction band and 0.21 eV/Å in the valence band for this calculation.[13,14] It should be noticed that the deformation potential value was taken from the value of InAlAs digital alloy as an approximation. $I(\nu,\nu';\boldsymbol{k},\boldsymbol{k}\pm\boldsymbol{q})$ is the overlap integral between the initial and final states, and $D_{\nu'}(E',\Omega_{\boldsymbol{k}'})$ is the density of states in $\Omega_{\boldsymbol{k}'}$, at energy $E' = E(k) \pm \hbar\omega_{\eta q}$. $\hbar\omega_{\eta q}$ is the phonon energy. Here, acoustic and optical phonon energies are taken to be the arithmetic average of the value for InSb (acoustic: 8 meV, optical: 25 meV) and AlAs (acoustic: 15 meV, optical: 50 meV).[15] $N_{\eta q} = \frac{1}{exp(\hbar\omega_{\eta q}/k_B T)-1}$ denotes the phonon occupation number. In this work, T=300 K.

The Keldysh model is used to calculate the ionization rate and expressed as

$$P_{\nu\nu'}(\boldsymbol{k},\boldsymbol{k}') = S(E-E_{th})^{\gamma}. \quad (2)$$

, where $E_{th}$ is the threshold energy, S is the softness parameter and $\gamma$ is an approaching index. The Keldysh parameters for electron and holes were adjusted by fitting the Gain curve and excess noise factor of Monte Carlo simulation results with experimental results. More detailed information about the Keldysh model (including how the parameters are fitted) can be found in Ref. 16,17. The parameters are provided in Table I.

TABLE I. Electron and hole parameters in the Keldysh model.

|  | S | $E_{th}$ | $\gamma$ |
|---|---|---|---|
| Electron | $2.1\times 10^{14}$ | 1.2 | 1 |
| Hole | $1.4\times 10^{15}$ | 1.4 | 1.1 |

In experiment, the epitaxial structure of the digital alloy APD from top to bottom is 100 nm p-GaSb ($1\times 10^{19}$ cm$^{-3}$), 100 nm p-Al$_{0.7}$InAsSb ($2\times 10^{18}$ cm$^{-3}$), 1000 nm un-doped Al$_{0.7}$InAsSb, 200 nm n- Al$_{0.7}$InAsSb ($2\times 10^{18}$ cm$^{-3}$). In the simulation, we take a simplification that the electric field distributes uniformly in the 1000 nm un-doped region and the depletion width is ignored. The simulated and measured gain curves are plotted in Figure 2(A) as dashed and solid curves, respectively. Figure 2(B) shows the simulated (◀) and measured (▶) excess noise factor versus gain. The simulation results are consistent with experimental results.[18]

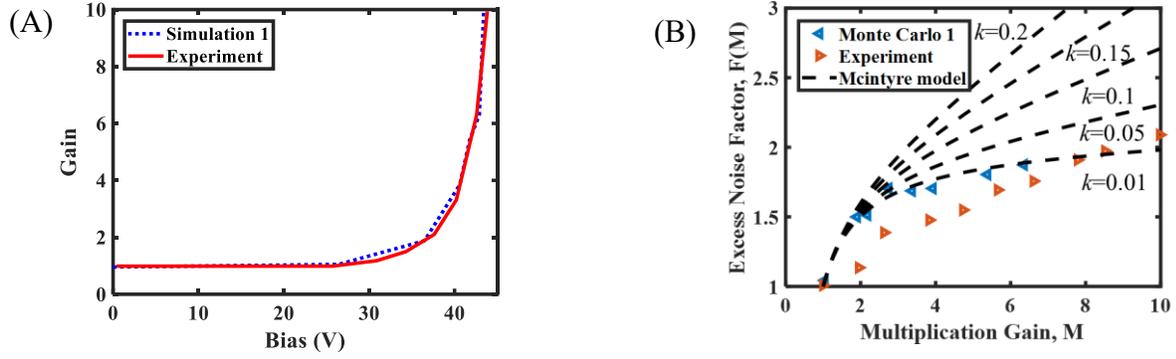

**FIGURE 2**  Full band-based Monte Carlo simulations for A, gain versus bias B, excess noise factor versus gain, M.

In our previous study, minigap in InAlAs digital alloy has been found to play an important role in influencing the ionization coefficient of holes.[13,14,19] The accelerating path of holes is blocked by minigap and hole energy is not enough to trigger ionizations. Thereafter, an extremely low $k$ value has been realized in InAlAs digital alloy. As shown in Figure 1(B), a large mingap has been found in the conduction band of AlInAsSb digital alloy, while, there is no significant minigap in the valence band. If the minigap in conduction band could block electrons like the case in the valence band of InAlAs digital alloy, we can infer that the ionization coefficient of electrons is lower than that of holes. However, we found in the experiment that the electron ionization coefficient is much larger than holes.[8] Apparently, minigap in the conduction band of AlInAsSb doesn't block the accelerating path of electrons, the electrons could find the path to jump across the minigap and gather enough energy to trigger ionizations. To verify this hypothesis, the transportation process of electrons is simulated based on full-band Monte Carlo simulation. Electrons that have an initial state above minigap and below minigap are analyzed and compared. As shown in Figure 1(B), electrons 1 and 2 are located at the Γ point (below minigap) and the first Brillouin zone boundary along [00-1] direction (above minigap), respectively. Electron 1 starts from the lower energy state and has to surmount the minigap to achieve sufficient energy to trigger ionizations. Electron 2 starts from a state above minigap and does not have the same energy barrier. The propagations of both electrons are tracked and the energies are recorded every 1 fs. The sampling points are over 1 million. According to the law of large numbers, if the sampling data is large enough, the energy distribution of one electron (or hole) can reflect that of all electrons (or holes). The energy distribution probability (normalized to the total number of sampling points) for the two types of electrons under an electric field of 439 kV/cm are shown in Figure 3. The electric field is taken considering the total bias is 43.9 V (avalanche gain is 10), avalanche region thickness is 1 μm and the electric field is uniformly distributed. There is little difference in the high-energy portion of the energy distributions, which indicates that the minigap does not prevent electrons from accumulating the energy required to impact ionize.

As was explained in our previous work[13], although the energy is discontinuous in the [001] direction, there are in-plane energy states that have sufficient energy to go across the minigap. The electrons that are scattered to these in-plane states can get over the minigap under the driving force of the electric field. In the simulation, the energy of electron 2 did not drop below the minigap because the 439 kV/cm electric field is high enough to enable electrons to accelerate to higher energy states. Also there are fewer energy states in the minigap energy level for electrons to relax to. The energies of holes are also shown in Figure 3. It can be seen that holes tend to occupy low energy states. The hole energies are smaller than the bandgap and do not trigger ionizations. It follows that the ionization coefficient of holes is much smaller than that of electrons.

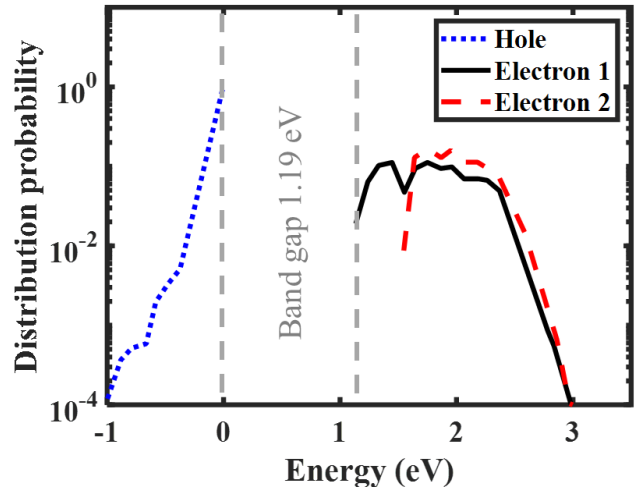

**FIGURE 3**  Energy distributions for carriers in AlInAsSb, wherein electrons with two types of initial states are compared.

In this work, 3-D band structure is used to deploy Monte Carlo simulation. To have an intuitive physical analysis of the energy distribution of carriers, some band structure parameters along different crystal directions are calculated and put in Table II. Important energy parameters are provided in Table III.

TABLE II. Effective masses for simplified single parabolic bandstructure

|  | Γ (001) | L (111) | X (100) | HH (001) | LH (001) | SO (001) |
|---|---|---|---|---|---|---|
| Effective mass ($m_0$) | 0.06 | 1.9 | 0.1 | 0.4 | 0.38 | 0.08 |

| | Bandgap | Split-off band energy gap | Energy step to the first mini-gap in the valence band | First Minigap in valence band | Acoustic phonon energy | Optical phonon energy |
|---|---|---|---|---|---|---|
| Energy ($eV$) | 1.19 | 0.48 | 0.66 | 0.2 | 0.0165 | 0.0364 |

TABLE III. Important energy parameter

Given the disparity between the electron mass and that of the hole in the heavy hole and light hole bands, it can be concluded that the split-off (*SO*) band plays a critical role in hole acceleration. Large separation (0.48 eV) between the heavy-light hole bands and the split-off band impedes impact ionization. The energy step to the first mini-gap is 0.66 eV, which is much smaller than bandgap (1.19 eV). Since the first mini-gap has an energy of 0.2 eV, which is much larger than phonon energy, the holes could not jump across the minigap by the help of phonon scatterings. Since the hole energy distribution is much more limited than that of the electron, we can infer that it is not easy for holes to find a path (in 3-D Brillouin zone) to get over minigap like electrons.

**IV. CONCLUSION**

In this paper, we present calculations of the band structure of digital alloy AlInAsSb. Avalanche photodiodes fabricated from AlInAsSb exhibit the lowest noise of any III-V compound material. By analyzing the band structure calculated by the empirical tight-binding method, we have shown that there are mini-gaps in both the conduction band and valence band. Full band structure based Monte Carlo simulation shows that minigap in the conduction band doesn't block the electron accelerating process. The hole accelerating process is limited, wherein large effective mass and minigap might play an important role. Good agreement between the full band Monte Carlo simulations and measured noise characteristics is demonstrated.

1. McIntyre RJ. Multiplication Noise in Uniform Avalanche Diodes. *IEEE Trans Electron Devices.* 1966;ED13(1):164-+.
2. Onabe K. Immiscibility Analysis for Iii-V Quaternary Solid-Solutions. *Nec Research & Development.* 1984(72):1-11.
3. Turner GW, Manfra MJ, Choi HK, Connors MK. MBE growth of high-power InAsSb/InAlAsSb quantum-well diode lasers emitting at 3.5 mu m. *Journal of Crystal Growth.* 1997;175:825-832.
4. Maddox SJ, March SD, Bank SR. Broadly Tunable AlInAsSb Digital Alloys Grown on GaSb. *Crystal Growth & Design.* 2016;16(7):3582-3586.
5. Tan Y, Povolotskyi M, Kubis T, Boykin TB, Klimeck G. Tight-binding analysis of Si and GaAs ultrathin bodies with subatomic wave-function resolution. *Phys Rev B.* 2015;92(8):085301.
6. Tan Y, Chen F, Ghosh A. First principles study and empirical parametrization of twisted bilayer MoS2 based on band-unfolding. *Appl Phys Lett.* 2016;109(10):101601.
7. Tan Y, Povolotskyi M, Kubis T, Boykin T, Klimeck G. Transferable tight-binding model for strained group IV and III-V materials and heterostructures. *Phys Rev B.* 2016;94(4):045311-045311.

8. Yuan Y, Zheng JY, Rockwell AK, March SD, Bank SR, Campbell JC. AlInAsSb Impact Ionization Coefficients. *IEEE Photon Tech Lett.* 2019;31(4):315-318.
9. Zheng J, Wang L, Wu X, et al. A PMT-like high gain avalanche photodiode based on GaN/AlN periodically stacked structure. *Appl Phys Lett.* 2016;109(24):241105
10. Zheng J, Wang L, Yang D, et al. Low-temperature-dependent property in an avalanche photodiode based on GaN/AlN periodically-stacked structure. *Sci Rep.* 2016;6.
11. Zheng J, Wang L, Wu X, et al. The Influence of Structure Parameter on GaN/AlN Periodically Stacked Structure Avalanche Photodiode. *IEEE Photon Tech Lett.* 2017;29(24):2187 - 2190.
12. Zheng JY, Wang L, Wu XZ, et al. Theoretical study on interfacial impact ionization in AlN/GaN periodically stacked structure. *Appl Phys Express.* 2017;10(7).
13. Zheng J, Yuan Y, Tan Y, et al. Digital Alloy InAlAs Avalanche Photodiodes. *J Lightwave Technol.* 2018;36(17):3580-3585.
14. Zheng J, Yuan Y, Tan Y, et al. Simulations for InAlAs digital alloy avalanche photodiodes. *Appl Phys Lett.* 2019;115(17).
15. Lindsay L, Broido DA, Reinecke TL. Ab initio thermal transport in compound semiconductors. *Phys Rev B.* 2013;87(16).
16. Keldysh LV. Kinetic theory of impact ionization in semiconductors. *Soviet Phys JETP.* 1960;37(10).
17. Osaka F, Mikawa T. Excess Noise Design of Inp/Gainasp/Gainas Avalanche Photodiodes. *IEEE J Quantum Electron.* 1986;22(3):471-478.
18. Yuan Y, Rockwell AK, Peng Y, et al. Comparison of Different Period Digital Alloy Al0.7InAsSb Avalanche Photodiodes. *J Lightwave Technol* 2019;37(14):3647-3654.
19. Zheng J, Tan Y, Yuan Y, Ghosh AW, Campbell JC. Strain effect on band structure of InAlAs digital alloy. *J Appl Phys.* 2019;125(8).